# Hard Magnetic Topological Semimetals in *X*Pt3: Harmony of Berry Curvature

Anastasios Markou[1][†], Jacob Gayles[2,1][*], Elena Derunova[3], Peter Swekis[1], Jonathan Noky[1], Liguo Zhang[1], Mazhar N. Ali[3], Yan Sun[1], Claudia Felser[1][†]

[1]*Max Planck Institute for Chemical Physics of Solids, 01187 Dresden, Germany*
[2]*Department of Physics, University of South Florida, Tampa, Florida 33620, USA*
[3]*Max-Planck Institute of Microstructure Physics, Weinberg 2, 06120 Halle (Saale), Germany*

Topological magnetic semimetals, like Co3Sn2S2 and Co2MnGa, are known to display exotic transport properties, such as large intrinsic anomalous (AHE) due to uncompensated Berry curvature. The highly symmetric *X*Pt3 compounds display anti-crossing gapped nodal lines, which are a driving mechanism in the intrinsic Berry curvature Hall effects. Uniquely, these compounds contain two sets of gapped nodal lines that harmoniously dominate the Berry curvature in this complex multiband system. We calculate a maximum AHE of 1965 S/cm in the CrPt3 by a state-of-the-art first principle electronic structure. We have grown high-quality thin films by magnetron sputtering and measured a robust AHE of 1750 S/cm for different sputtering growth conditions. Additionally, the cubic films display a hard magnetic axis along [111] direction. The facile and scalable fabrication of these materials is prime candidates for integration into topological devices.

*gayles@usf.edu
†markou@cpfs.mpg.de, felser@cpfs.mpg.de

The nontrivial topology of the electronic materials has come to the forefront of research to characterize and to predict observable phenomena of exciting materials[1]. These phenomena are theoretically realized through the adiabatic transport of quasiparticles known as the Berry curvature[2]. The Berry curvature acts as an emergent electromagnetic field on the wavepackets, which adds an additional phase. For spinful electron wavepackets, this may result in such real observables as the anomalous and spin Hall effects (AHE & SHE). Both effects, able to convert charge currents to spin currents[3,4], manifest as a transverse motion of the to an applied current, that originates from extrinsic and intrinsic mechanisms due to the electronic structure[5]. These phenomena also have thermoelectric counterparts when a temperature gradient replaces the applied current and are correspondingly termed Nernst effects[2,6]. Specifically, the intrinsic mechanism is tied directly to the electronic structure to produce the Berry curvature, which is determined by the underlying crystal symmetry. From this, one can uniquely predict the type and size of the Berry curvature in any crystalline material solely based on the atomic constituents and crystal symmetry. Recently, topological magnetic semimetal materials, $Co_3Sn_2S_2$[7–9] and $Co_2MnGa$[10,11], in addition to others like $Fe_3GeTe_2$[12], and $Fe_3Sn_2$[13], have attracted great interest due to the topological nature of the electronic structure. $Co_3Sn_2S_2$ and $Co_2MnGa$ both host Weyl crossings near the Fermi energy between two topologically connected electronic bands[14]. In the absence of the relativistic spin-orbit coupling (SOC), calculations show a nodal line/ring, where the two bands are degenerate in momentum space. The addition of SOC gaps the nodal line at all momenta except to leave the two Weyl nodes, which act as a source and sink of Berry curvature[15]. This Berry curvature results in an AHE due to the lack of time-reversal symmetry that is sizable and experimentally confirmed at room temperature[16,17]. A realization that has led to the possibility for quantum anomalous Hall effect in confined dimensions[18], pristine topological surface states[9,11,19], and highly conductive metals[7,20]. The goal then is to generate large sources of uncompensated Berry curvature between bands by tuning the SOC strength, symmetry, and dispersion of the bands.

The Berry curvature of electronic band structures determines the intrinsic AHE. Band anti-crossings with spin splitting lead to a robust local Berry curvature; therefore, it can be used as a guiding principle to design strong intrinsic AHE. The effective overlap between band anti-crossings at the Fermi level is required to have strong net Berry curvature, such as nodal lines. So far, most studies focus on the design of strong intrinsic AHE via anti-crossings between two crucial bands, such as the two degenerate bands constructing nodal lines. In this work, we propose a new strategy by considering both the band anti-crossing band structures and the sign of the generated

Berry curvatures. The AHE can be strongly enhanced in the case of multifold band ant-crossings making harmonic contributions, as illustrated in Fig 1b. This AHE is realized in the highly symmetric cubic $X$Pt$_3$ compounds, in Fig 1a, with our experimental sputtered thin films that display a hard magnetic phase, an agreeable AHE that survives above room temperature. The $X$Pt$_3$ compounds host some of the largest magneto-optical Kerr effects[21–24], an effect that is the AC equivalent of the AHE Berry curvature. We focus on the electronic structure in compounds $X$=V, Cr, and Mn with alloying of the transition metal $3d$ atoms. We thereby tune the magnetic exchange field, the SOC, and electron occupation to detail the robustness of the observed phenomena. We also grow high-quality thin films of CrPt$_3$ and see a significant contribution to the intrinsic Berry curvature AHE even at room temperature.

In Fig. 1 we explore the electronic structure properties of the $X$Pt$_3$ compounds. Fig. 1 shows the crystal structure of $X$Pt$_3$ $Pm\overline{3}m$ cubic compounds with the $X$ ion (blue) sitting at the corners and the Pt ion (silver) on the faces. The compound is highly symmetric. However, when magnetization is considered, the symmetry reduces to a tetragonal magnetic symmetry $P4/mm'm'$. When the magnetization is allowed along the c-axis, the symmetry of the Pt ions on the top/bottom faces (Pt $1c$) differentiates from the side faces (Pt $2e$). In Fig. 1 b we display a schematic of the enhancement of the Berry curvature. In figures 2 a-c, we plot the electronic band structures along with the Fermi energy-dependent AHC in the parent $X$Pt$_3$ compounds. In Fig 2 a,b, and c, we plot the electronic band structure in the left panel along with the AHC in the right panel of VPt$_3$, CrPt$_3$ and MnPt$_3$, respectively. For all plots, we see a maximum AHC of nearly 2600 S/cm, which shifts from just above the Fermi level in VPt$_3$ to just below the Fermi level in MnPt$_3$. The shape of the maximum is nearly consistent across all samples, which we attribute to two sets of two bands. These sets of bands are colored red and blue in the single particle electronic structures. The orbital bands separate for $d$ orbitals along the z-axis, $d_{xz}, d_{yz}, d_{z^2}$ and those in the xy-plane $d_{xy}$, and $d_{x^2-y^2}$. Due to the tetragonal magnetic symmetry, the Pt at $2e$ (side face) shows distinct DOS compared to the Pt $1c$ on the (top/bottom faces), which is included in the supplementary. Gapped regions between two eigenvalues allow for the finite Berry curvature,

$$\Omega_{ij}^n = \sum_{m \neq n} \frac{<n|\hat{v}_i|m><m|\hat{v}_j|n>}{(E_n - E_m)^2} \quad \text{Eq. 1}$$

, whose summation over the entire Brillouin zone is equivalent to Eq. 2. In Eq. 1, $\hat{v}_i$ is the velocity operator for an $i$ direction. Correspondingly, we plot the summed Berry curvature of the two sets of bands against the total AHC. The four bands below and above the highlighted bands are also plotted against (dashed grey) the total AHC. The eigenvalue of the velocity operator is dependent on the orbital character of the band.

In Fig. 3a, we compare magnetization, AHC, and anomalous Nernst conductivity for various alloying strengths of Cr into V and Mn. The top panel of Fig 3a details the magnetic moment of the $X$ ion in black that ranges from 1.5 to 3.7 $\mu_B$ on the left y-axis. On the right y-axis, we depict the orbital moment of the $X$ ion in dashed blue and the average magnetic moment of the Pt ions in red. The Pt moments are an order of magnitude smaller than those of the $X$ ions, as expected. The orbital moment is largest for Cr due to the hybridization with the Pt orbitals. When the magnetization is allowed along the c-axis, the symmetry of the Pt ions on the top/bottom faces is differentiated from the side faces, which trends to be ten percent larger in the VPt$_3$ to ten percent smaller in the MnPt$_3$. Furthermore, the average moment of the Pt ion changes sign with an increase in Cr content, which aligns antiparallel in V saturated alloys and parallel in the Mn saturated (top panel Fig 2a). From the $X$Pt$_3$ compounds, Pt has the largest source of SOC, with a small increase of SOC from V to Cr to Pt. SOC lifts the orbital degeneracy, which is crucial for transport phenomena such as the AHC and the temperature gradient analog the anomalous Nernst effect (ANC). In the bottom panel of Fig 3 a, we plot the AHC in blue on the left y-axis and the ANC on the right y-axis as a function of the $X$ ion content. The SOC causes nontrivial band topology to arise when the conduction band inverts with respect to the valence band. In momentum space, this causes a finite Berry curvature to accelerate the spin quasiparticles transversely. The Kubo formula expresses a linear response calculation of the AHC from the Berry curvature for a 3D system:

$$\sigma_{ij} = \frac{e^2}{\hbar} \sum_n \int \frac{d^3k}{(2\pi)^3} \Omega_{ij}^n f(\varepsilon_k) \qquad \text{Eq. 2}$$

$f(\varepsilon_k)$ is the Fermi level, and $\Omega_{ij}^n$ is the Berry curvature of the n'th band, which is integrated over all momentum space. The ANC is calculated on from the same Berry curvature as an expansion of the Mott relation for large temperatures[2,26]. In the bottom panel, we see that the AHC ranges from 1000 to 2500 S/cm and is largest for an equal alloying of V and Cr. Whereas the ANC, interestingly ranges from -7 to 4 A/mK and changes sign twice as a function of the $X$ ion alloying. Additionally, in Fig 3, we explore the degenerate spin bands without SOC and the Berry curvature with SOC of CrPt$_3$. In Fig 3 b and c, the spin-polarized bands without SOC show degenerate bands in the [100] and [110] planes. The color distribution shows the difference between two spin-polarized bands in the plane. These degenerate lines in blue/red that are nodal lines are shown for spin-polarized bands that correspond to the SOC bands of Fig 2 b. In Fig 3 d we show the momentum distribution of the Berry curvature in 3D with SOC. In Fig 3 e and f, we plot the Berry curvature of all occupied bands on the [100] and [110] Brillouin zone face.

Along the [001] direction, the crystal mirror planes $m_x$ and $m_y$ differ from $m_z$. Similarly, the mirror planes along the diagonal $m_{xy}$ differ from the mirror planes that have a normal component that is parallel to the tetragonal axis. SOC causes the nodal lines to gap in the Brillouin zone that has only the $m_x$, $m_y$, and $m_{xy}$. The distribution of the Berry curvature is centralized around the nodal lines, which are gapped by SOC. Furthermore, there is a large shift of the bands from the non-SOC to the SOC case.

To verify our theoretical predictions, we grew a high-quality cubic $L1_2$-CrPt$_3$ film by magneto-sputtering. Figure 4 summarizes the structural and magnetic of the 60 nm thick CrPt$_3$ films. We grew two distinct films with Cr$_{22}$Pt$_{78}$ and Cr$_{24}$Pt$_{76}$ of the same thickness to observe the robustness of our predictions. Fig. 4 a depicts a symmetric $\theta$-$2\theta$ x-ray diffraction (XRD) scan of the Cr$_{22}$Pt$_{78}$ in the top panel (red) and Cr$_{24}$Pt$_{76}$ in the bottom panel (blue) film grown on (0001) Al$_2$O$_3$ substrate. We exclusively observe the (111) family reflections of the CrPt$_3$, which implies the films are (111)-oriented and single-crystalline. The (111)-oriented growth ensures a cubic structure with equivalent strain on the three crystal axes. In Fig. 4 b, we plot the in-plane (IP) and out-of-plane (OOP) magnetization hysteresis loops at 100 K of Cr$_{22}$Pt$_{78}$ in the top panel and Cr$_{24}$Pt$_{76}$ in the bottom panel. The $L1_2$-type (111) oriented films exhibit perpendicular magnetic anisotropy (PMA) along the [111] direction. The easy magnetization axis is normal to the film plane along the [111] direction. The magnetization of the sample directly affects transport and especially Hall transport. We patterned our film into Hall bar devices to perform magnetotransport measurements. The lattice constants of Cr$_{22}$Pt$_{78}$ and Cr$_{24}$Pt$_{76}$ are 3.784 and 3.782 Å, respectively. Figure 5 a shows the longitudinal resistivity ($\rho_{xx}$) as a function of temperature ($T$) measured at zero magnetic fields of the Cr$_{22}$Pt$_{78}$ in the top panel and Cr$_{24}$Pt$_{76}$ in the bottom panel. The samples are both metallic and show an RRR ratio of 1.7 and 1.8. The Cr$_{24}$Pt$_{76}$ starts at a low temperature, starts with lower resistivity than Cr$_{22}$Pt$_{78}$, and at high temperatures surpasses the Cr$_{22}$Pt$_{78}$. The hysteresis is also seen in magnetoresistance ratio (MR), $\frac{\rho_{xx(H)} - \rho_{xx(H=0)}}{\rho_{xx(H=0)}}$, shown in Fig 5 b as a function of temperature. In Fig 5 c, we plot the magnetoresistance as a function of the external magnetic field direction with respect to the in-plane current. We have extracted the mobilities to be 2.8 $\frac{cm^2}{V \cdot s}$ at 2K, which is comparably larger than other sputtered films, as detailed in the supplementary.

In Fig. 6 a we plot the Hall resistivity (subtracted ordinary Hall resistivity) as a function of external OOP magnetic sweeps at temperatures 2, 100, 200, and 300 K for both concentrations. The compounds both show a hysteresis of the Hall resistivity with the direction of the sweeping field that follows the hysteresis of the magnetization. Phenomenologically the anomalous Hall resistivity, $\rho_{xy}$, scales with the longitudinal resistivity, $\rho_{xx}$, through the extrinsic side-jump and skew scattering mechanisms. The intrinsic, Berry curvature contribution can be determined by fitting the following equation:

$$\frac{\rho_{xy}}{\mu_0 M \rho_{xx}} = \alpha + (\beta + b)\rho_{xx}, \qquad \text{Eq. 3}$$

where $\mu_0$ is the magnetic permeability and $M$ is the magnetization, and $b$ is determined by the electronic structure that gives rise to the Berry curvature detailed in Fig 1, and 2. $\beta$ is the extrinsic side-jump constant, and $\alpha$ is the skew scattering constant. In Fig. 6 b we fit both concentrations of the CrPt3 by the variation of $\rho_{xx}$ through temperature from 50 K to 200 K. From Eq. 3, we can separate the skew scattering from $(\beta + b)$. The dashed black line in Fig. 6 b shows the quality of the fit of the experimental points. The colored solid line is the impurity density-independent ($\beta + b$) contributions, and the dashed colored line is the skew scattering contribution ($\alpha$). For both concentrations, the linear scaling of the impurity density-independent terms appears to be quite large in the order of the theoretically calculated value of 1733 S/cm for the [111] direction with perfect stoichiometry. Whereas, in the Cr22Pt78, the $\alpha$ is nearly zero compared to the more considerable value seen in Cr24Pt76. To further understand the underlying mechanisms of CrPt3. In Fig 6 c we plot the longitudinal conductivity, $\sigma_{xx}$, versus the anomalous Hall conductivity, $\sigma_{xy}$, on a log-log plot for both CrPt3 concentrations, well-known elemental ferromagnets, and the magnetic Weyl semimetals in literature. The vertical dashed lines show the expected boundaries for the intrinsic metallic regime. Bellow, the first vertical line, $\sim\sigma_{xx} < 10^3$, is the localized hopping regime, and above, $\sim\sigma_{xx} > 10^6$, is the skew scattering regime. The red horizontal line shows the theoretical value of 1200 S/cm for Co3Sn2S2. The theoretical value of CrPt3 is shown as the dashed blue line.

The most striking feature of this work is the conclusive plot in Fig 6 c. We see that typically the elemental and many ordinary metals have an intrinsic AHC that lies just below this high conductivity region of $> 10^6$. The high conductivity materials typically show relatively low intrinsic AHC but can become extremely large in the high conductivity regime for topological metals[27]. Whereas the magnetic Weyl semimetals display an AHC that lies just above the region of $< 10^3$. In $Co_3Sn_2S_2$ and $Co_2MnGa$, the single set of linear topological crossings have a minuscule DOS with comparatively low conductivity but have shown some of the largest intrinsic AHCs. Whereas the CrPt$_3$ not spontaneous but remarkably lies directly in the middle of these two regions of $\sigma_{xx}$ with a clear intrinsic AHC that reaches the upper echelons of known literature values. This is unique to semimetals, due to the continuous gapping of bands except at the degenerate Weyl points. Furthermore, in the *X*Pt$_3$ compounds show a larger dispersion where both sets of topological bands can have a contribution at the Fermi energy. Commonly, in many magnetic metals, there are multiple sets of bands, with the exception that they display a compensating Berry Curvature. The highly cubic crystal symmetry allows for the nodal lines, and the SOC supplied by the Pt gaps the nodal lines. And the hybridization of the Pt with the *X* ion allows for this harmonious interaction of topological bands.

The thin-films grown are of high-quality and further point to the robustness of the predicted and observed effect. Firstly, both films produced with a variance of the skew scattering contribution distinctively display a sizable AHC in comparison to other grown films. The MR of the two films is comparable, with an increase in MR when the external field and current are parallel. When compared to other sputtered topological metals, the CPt$_3$ displays a pristine quality and a distinct intrinsic AHC. When combined with the unconventional hard magnetic properties in a cubic magnet, and high Curie temperature, the CrPt$_3$ class of materials are ideal for further research and integration into device applications.

In summary, we predicted from bandstructure analysis that the sizeable transverse magnetotransport in *X*Pt$_3$ alloys is due to the combination of two gapped nodal lines that synergistically work to generate Berry curvature. We have shown experimentally that this mechanism is robust in thin films on 60 nm thick, which enters a new regime for the AHE of thin films. The robustness of the band structures may allow for the study of topological materials in the 2D limit[18]. These results allow for an approach to increase the size of the Berry curvature systematically. Where Heusler materials, specifically the ternary and quaternary, are an interest due to their high tunability and cubic nature, that allow mirror planes. Such quantum materials can lead the path to engineering next-generation devices.

**METHODS:** We carried out state-of-the-art density functional theory calculations using the full potential linearized augmented plane wave method of FLEUR[25] within the generalized gradient approximation (GGA) exchange-correlation functional for bulk alloys of $X$Pt$_3$. We calculated the magnetic properties of the alloy materials using the virtual crystal approximation along with the Vegard's lattice constant from the experimental lattice constants of pure VPt$_3$, CrPt$_3$, and MnPt$_3$. We converged our collinear calculations with a plane wave cutoff of 3.9 a.u.$^{-1}$ and $20^3$ **k**-points in the full Brillouin zone. For the calculation of the AHE Berry curvature, we used 72 Wannier functions, projected on to $s$, $p$, and $d$ orbitals for each atom, with spin-orbit coupling for each unit cell. The calculations of the anomalous Hall conductivities are converged on a $512^3$ **k**-point interpolated grid in the full Brillouin zone.

The 60 nm thick CrPt$_3$ films were grown using magnetron sputtering in a BESTEC ultrahigh vacuum system (UHV) with a base pressure less than $2 \times 10^{-9}$ mbar and a process gas (Ar 5 N) pressure of $3 \times 10^{-3}$ mbar. We fix the target to substrate distance to 20 cm, and the substrate was rotated during deposition to ensure homogeneous growth with a speed of 20 rpm. The CrPt$_3$ Co-sputtered films were grown on $10 \times 10$ mm single-crystal $c$-cut (0001) Al$_2$O$_3$ substrate from Cr (5.08 cm) and Pt (5.08 cm) sources in confocal geometry, using 25 W, and 38 or 42 W DC power, respectively. The films were grown at 800 °C and then post annealed *in situ* under UHV at 850 °C for 60 min to improve the crystallinity. Finally, a 4 nm Si capping layer was deposited at room temperature using a Si (5.08 cm) target at 60 W RF power to prevent oxidation of the CrPt$_3$ epilayer. The growth rates and the film thickness were determined by a quartz crystal microbalance and confirmed by using XRR measurements.

XRD and XRR were measured in a Panalytical X'Pert3 MRD diffractometer using Cu K$\alpha_1$ radiation ($\lambda = 1.5406$ Å). Stoichiometry was estimated as Cr$_{22}$Pt$_{78}$ and Cr$_{24}$Pt$_{76}$ at.% by energy-dispersive x-ray spectroscopy (EDXS) with an experimental uncertainty of 2%. The sample's magnetization was carried out using a Quantum Design MPMS3. We performed electrical transport measurements in Hall bar devices with a size of 250 × 50 µm, patterned using a combination of optical lithography and Ar ion etching, set up in an eight-terminal geometry using Al wire-bonded contacts. Magnetotransport was measured using a low-frequency alternating excitation current (ETO, PPMS9 Quantum Design). The electrical current used for both longitudinal and transverse resistance measurements was 100 µA.

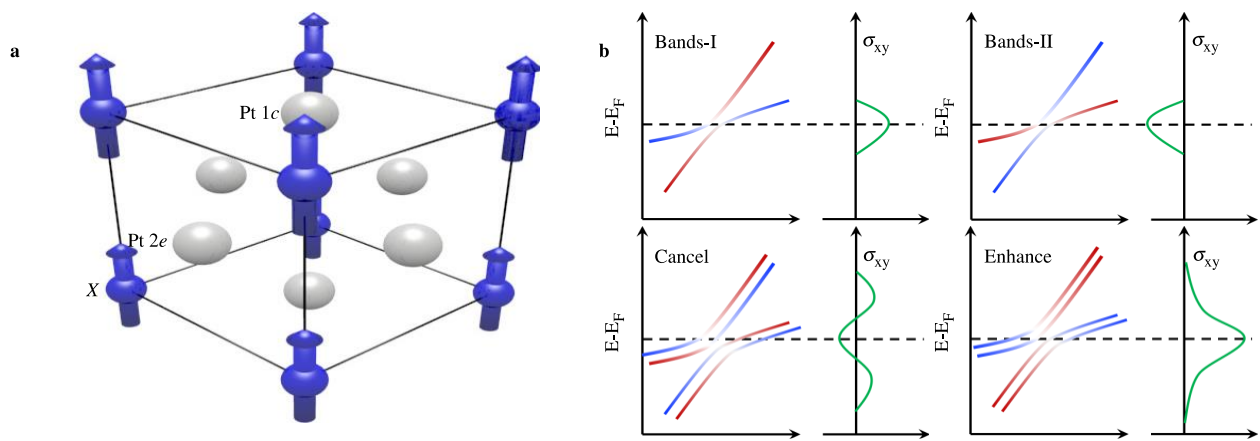

**Fig. 1. Electronic structure and transport. a** The crystal structure of $X$Pt$_3$. **b** Schematic of multiple sets of bands interactions.

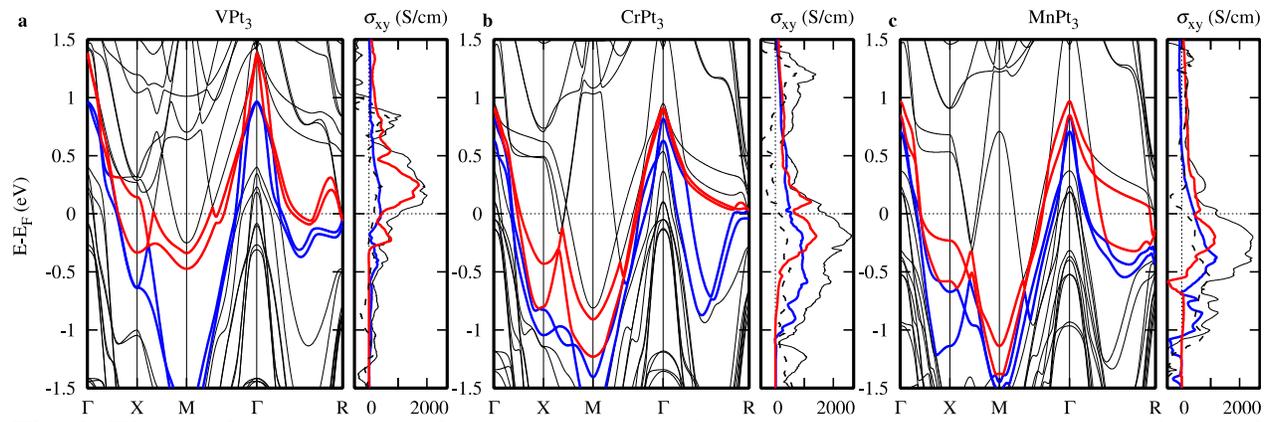

**Fig. 2. Electronic structure and transport. a** Band structure and Fermi energy dependence of AHC in VPt$_3$ along with band projections. **b** Band structure and Fermi energy dependence of AHC in CrPt$_3$ along with band projections. **c** Band structure and Fermi energy dependence of AHC in MnPt$_3$ along with band projections.

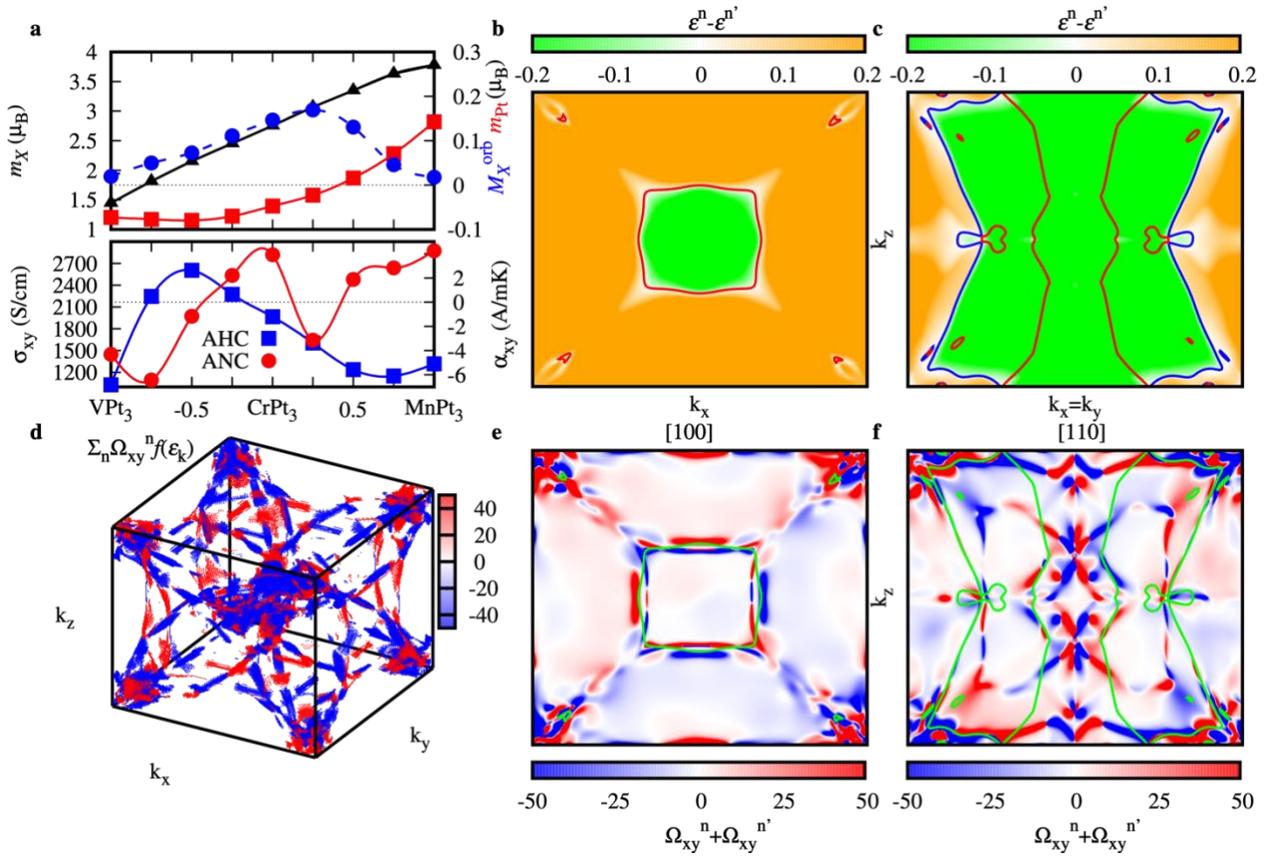

**Fig. 3. Band analysis and Berry curvature. a** top panel: Magnetic moment and orbital moment $X$ ion in black and blue respectively, and the magnetic moment of Pt in red as a function of $X$ ion concentration. **b & c** Spin-polarized band difference and nodal line of spin bands in the [100] and [110] plane respectively. **d** 3D distribution of Berry curvature at the Fermi energy in the irreducible Brillouin zone. **d & f** 2D distribution of Berry curvature of gapped bands from the Nodal line on the [100] and [110] plane, respectively.

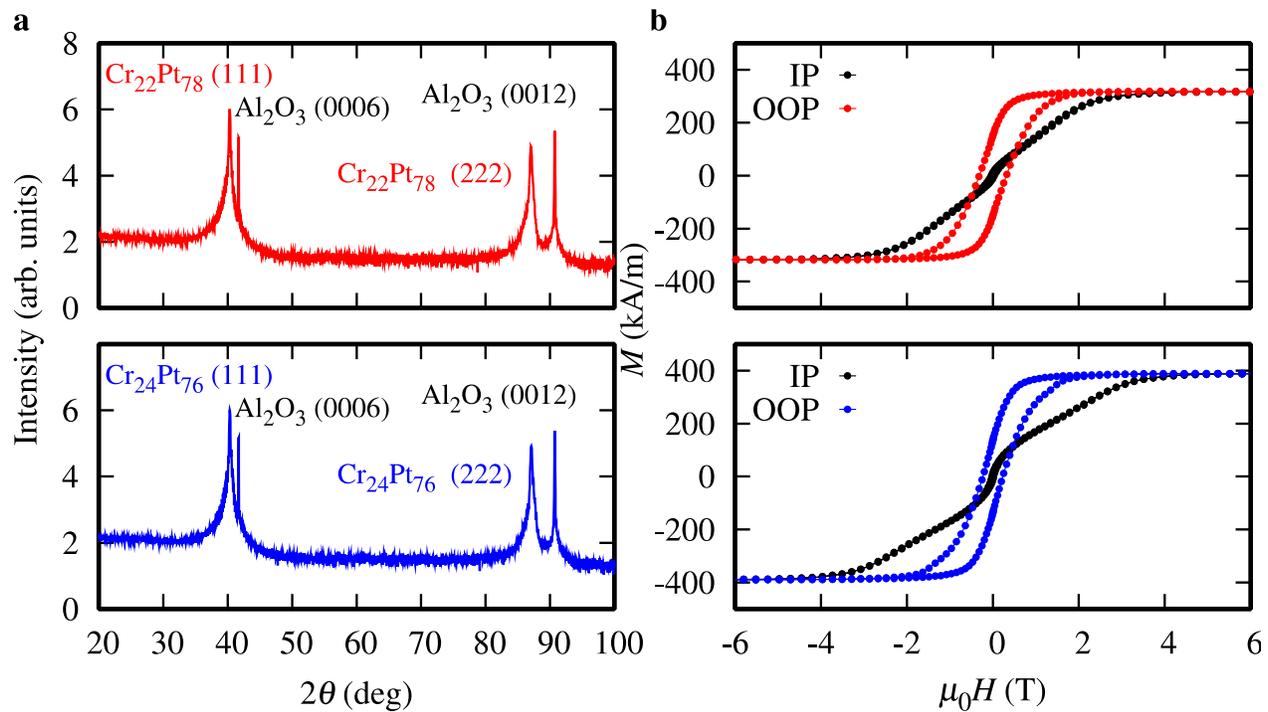

**Fig. 4. Structural, magnetic, and electrical properties. a** XRD pattern of a 60-nm-thick CrPt$_3$ film on a Al$_2$O$_3$ (0001) substrate. **b** Magnetization hysteresis loops measured at 100 K with the applied field in-plane along the [1$\bar{1}$0] direction and out-of-plane along the [111] direction.

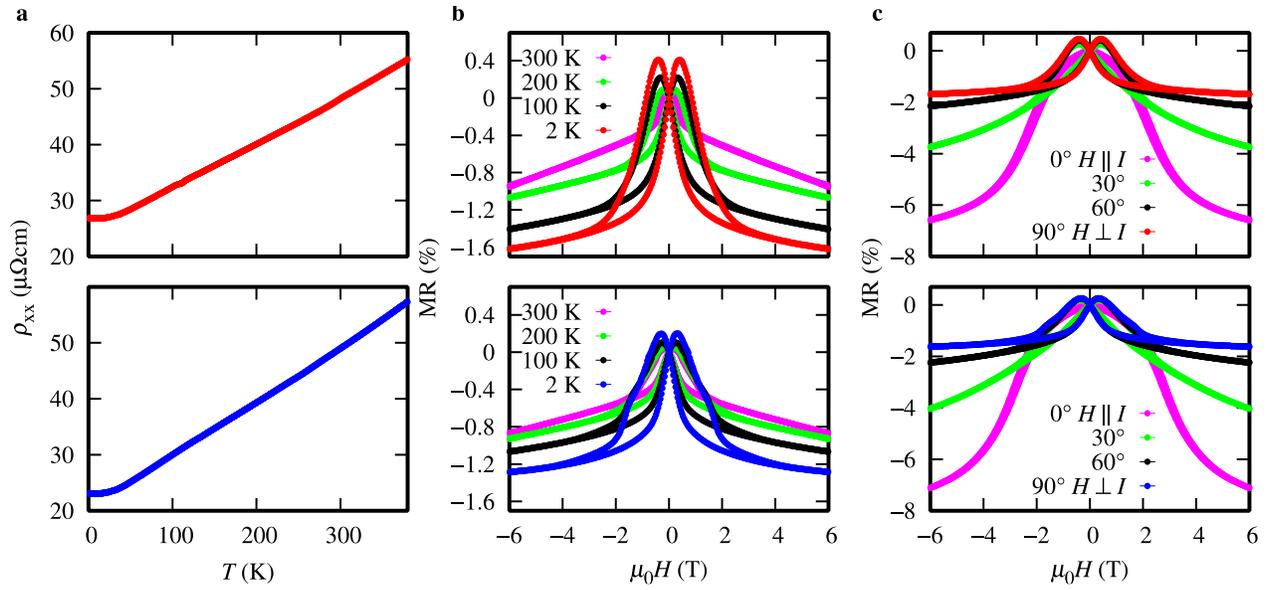

**Fig. 5. Structural, magnetic, and electrical properties. a** Longitudinal resistivity measured as a function of temperature on zero-field cooling for a $250 \times 50~\mu m^2$ Hall bar (inset). **b** Magnetoresistance as a function of external field and temperature. **c** Magnetoresistance as a function of the angle between the external field and current direction.

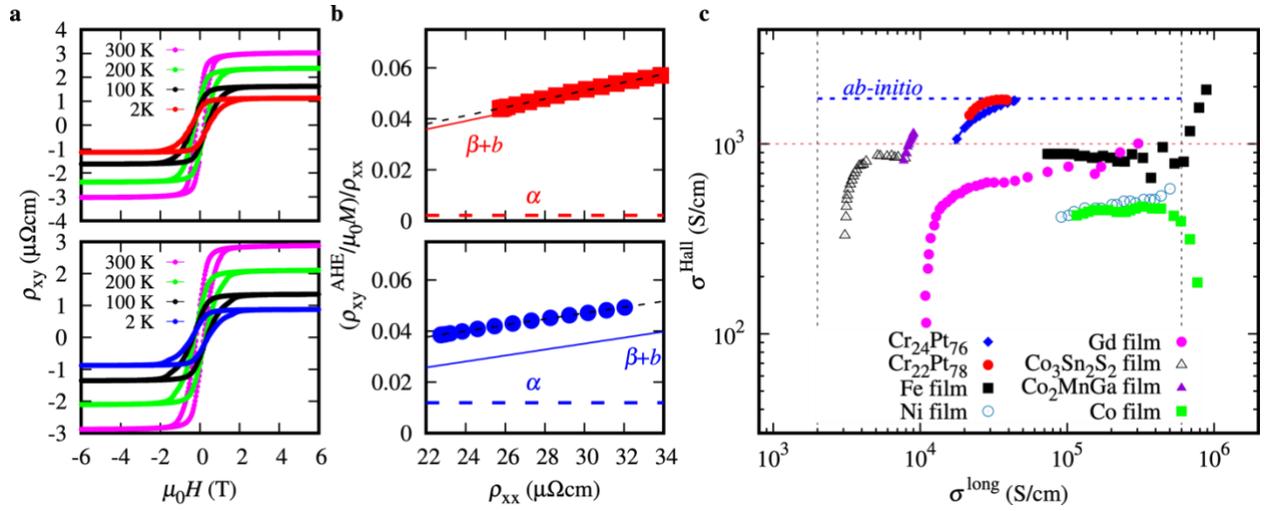

**Fig. 6. Anomalous Hall effect. a** Hall resistivity as a function of external field and temperature. **b** Extraction of skew scattering ($\alpha$ dashed lines) and the side-jump+intrinsic ($\beta + b$ solid lines) from the experimental Cr$_{22}$Pt$_{78}$ (red) and Cr$_{24}$Pt$_{76}$ (blue). **c** Log-log plot of longitudinal conductivity versus the Anomalous Hall conductivity and compared to well-known metallic and semi-metallic films.

*ACKNOWLEDGMENTS:*

The authors acknowledge the innovation program under the FET-Proactive Grant agreement No. 824123 (SKYTOP). This work was financially supported by the ERC Advanced Grant No. 742068 "TOPMAT". We also gratefully acknowledge Max Planck Computing and Data Facility (Garching, Germany) for providing computational resources.